# State-Dependent Electron Delocalization Dynamics at the Solute-Solvent Interface: Soft X-ray Absorption Spectroscopy and Ab Initio Calculations

Sergey I. Bokarev[1], Marcus Dantz[2,3], Edlira Suljoti[2,3], Oliver Kühn*[1], Emad F. Aziz*[2,3]

[1]Institut für Physik, Universität Rostock, Universitätsplatz 3, D-18055 Rostock, Germany
[2]Joint Ultrafast Dynamics Lab in Solutions and at Interfaces (JULiq), Helmholtz-Zentrum Berlin für Materialien und Energie, Albert-Einstein-Strasse 15, D-12489 Berlin, Germany
[3]Freie Universität Berlin, FB Physik, Arnimallee 14, D-14195 Berlin, Germany

Non-radiative decay channels in the L-edge fluorescence yield spectra from transition metal-aqueous solutions give rise to spectral distortions with respect to X-ray transmission spectra. Their origin is unraveled here using partial and inverse partial fluorescence yields on the micro-jet combined with multi-reference *ab initio* electronic structure calculations. Comparing $Fe^{2+}$, $Fe^{3+}$, and $Co^{2+}$ systems we demonstrate and quantify unequivocally the state-dependent electron delocalization within the manifold of d-orbitals as one origin of this observation.

* Corresponding authors: Emad.Aziz@fu-berlin.de, Oliver.Kuehn@uni-rostock.de

*PACS: 31.15.A, 31.15.vj, 31.70.Dk, 32.80.Aa, 78.70.En*

Interfaces between solvents and ionic species, where bond-making and -breaking takes place via valence molecular orbitals (MOs), are key to the function of materials from the molecular up to the nanoscale and to the understanding of different chemical and biological processes [1, 2]. Water is one of the major hosting solvents and plays a variety of roles at different levels of complexity, from molecules and cells to tissues and organisms [3-8]. A direct way to probe the local structure of this interface is to investigate the nature of the involved unoccupied MOs via atom specific core-level X-ray absorption (XA) spectroscopy [9, 10]. In case of TMs, excitation from and relaxation to the L-edge addresses directly the empty valence and occupied d-MOs, respectively. Here a new mechanism has been presented recently based on the observation of dips/peak-reduction in the total fluorescence yield (TFY). These observations had been interpreted based on the electronic structure of the solute-solvent interface region [11, 12]. This observation triggered a debate in the field of X-ray optics, because in principle these dips could be attributed to two effects: First, a change of the ratio of solvent background (i.e. from oxygen atoms) to the solute's metal L-edge emission [13-15] and second, a charge delocalization between the metal d-orbitals and the p-orbitals of the solvent, which could be explained by electronic configuration mixing between ions and water molecules [15, 16]. This would correspond to an ultrafast electron relaxation, with the electron dynamics occurring within the core-hole lifetime of a few femtoseconds. Achkar *et al.* and de Groot argued that only the first interpretation is reasonable [14, 17], whereas Aziz *et al.*, on the basis of further investigations [12, 16], concluded that both effects are co-existing [15]. It should be pointed out that the TFY reported in earlier studies [11, 12] is not conclusive in this respect since it does not provide direct and quantitative information on the electronic relaxation pathways of the 2p-core excited TM ion [10, 18]. The partial fluorescence yield (PFY) measured for the metal center is more sensitive and avoids the solvent background [19]. Furthermore, the inverse partial fluorescence yield (iPFY) can be measured by inversion of the signal due to non-resonantly excited solvent atoms in the presence of the metal. This method for measuring XA spectra has also been proven to be bulk sensitive and able to give absorption spectra free from optical artifacts (e.g. saturation effect) [17, 20, 21]. Wernet *et al.* performed PFY measurements and argued that only the first interpretation (called optical effect) is valid [22]. However, their argument has been mostly based on semiempirical simulations of spectral lineshapes. Furthermore, Lange *et al.* [23] have observed a reduction of bands at the $L_3$-edge of iron in the active centre of myoglobin upon ligation with different ligands, which cannot be attributed solely to the optical effect. In that study concentrations were kept constant far below the level of the optical effect discussed in Ref. [22].

For a quantitative interpretation of iPFY spectra a first principles model is required, which can account for effects like solvent-induced electron-delocalization. Ligand field multiplet theory [24], widely used for interpretation of experimental results [10], being a semiempirical approach and treating the metal center in the field of the ligand's point charges, is not suitable for this purpose. Focusing on MO calculations of core-excitation spectra, (time-dependent) density functional theory (TDDFT) and non-empirical multiplet DFT based configuration interaction methods, allowing for the treatment of metal ions in their environment explicitly, were successfully applied for predicting K-edge [25-29] and L-edge [30-33] XA spectra. But, in contrast to K-edge spectra, MO investigations of L-edge spectra are not well established. Here, an important point concerns the electronic multi-configurational nature of the ground and excited states of TM complexes, which cannot be treated accurately by single-reference DFT methods.

The present work combines the power of the recently introduced iPFY and PFY measurements on a micro-jet in the soft X-ray regime, with a first principles based assignment. This allows drawing a comprehensive picture of the nature of MOs under the L-edge of TMs and their role in electron delocalization. Specifically, the multi-reference Restricted Active Space Self-Consistent (RASSCF) method together with a State-Interaction treatment of Spin-Orbit effects (RASSI-SO) is used to calculate L-edge spectra of aqueous metal ions. On this basis a simple two-state model is formulated to explain the state-dependent electron delocalization dynamics, which leaves its fingerprints in fluorescence yield spectra. For illustration and validation of the method, different ions and oxidation states ($Fe^{3+}$, $Fe^{2+}$, and $Co^{2+}$) are considered that lead to specific spectral features and interpretations. The former two systems have been the subject of Refs. [11-16, 22, 34]. We have used 1M concentration to compare with the previous analysis where the pH was close to neutral (~6) for $Fe^{2+}$ and $Co^{2+}$, and 0.5 for $Fe^{3+}$.

Spectra were obtained using the newly developed high-resolution X-ray emission spectrometer with a Rowland-circle and a liquid micro-jet as presented before in Refs. [19, 20] (cf. Fig. S1a (Supplementary Material)). The PFY spectra were recorded by setting the spectrometer to a given emission line and scanning the incident energy in the region of the TM $L_{2,3}$ absorption edges ($Co^{2+}$, $Fe^{2+}$, and $Fe^{3+}$). For the iPFY, inversion of the non-resonantly excited oxygen K-edge fluorescence is used [19].

In the theoretical simulations, $[M(H_2O)_6]^{n+}$ ions (M = $Co^{2+}$, $Fe^{2+}$, and $Fe^{3+}$) as well as $[FeCl(H_2O)_5]^{2+}$ and $[FeCl_2(H_2O)_4]^+$ were calculated on the RASSCF level with the relativistic ANO-RCC-VTZ basis set [35, 36]. The RASSI-SO treatment [37] included directly interacting states with $\Delta S = 0, \pm 1$. RASSCF/RASSI calculations were performed with MOLCAS 7.6 [38] (for details, see Supplementary Material).

In Figure 1, experimental TFY, PFY and iPFY spectra for the $Fe^{3+}$ ion in water are presented. In comparison to the iPFY spectrum, which represents the absorption cross section free of artifacts [17, 20, 21], the PFY shows a lower pre-peak intensity (at ~708 eV) and even a dip in case of the TFY, and a higher intensity for the $L_2$-peak (721.0-726.5 eV). An analogous increase of $L_2$ intensity in the PFY, if compared to iPFY, is shown in Figure 2 for $Fe^{2+}$ and can be attributed to the Coster-Kronig effect for both ions [19]. The remaining parts of the spectra are similar for TFY, PFY and iPFY. Figure 3 shows the L-edge transmission and TFY XA spectrum of a $CoCl_2$ aqueous solution. Here we use the transmission XA spectrum as a reference for the true absorption cross section, thus proving that iPFY indeed is a method capable of measuring absorption free of artifacts [16, 20]. The Co $L_{\eta,l}$ PFY (corresponding to 3s→2p transitions) is in good agreement with the transmission spectrum after area normalization. Since it depends on the relaxation of the 3s-electron to the 2p core-hole, it gives unquenched photon-out events. Interestingly, the pre-peak (~775 eV) in the Co $L_{\alpha,\beta}$ PFY spectrum (3d→2p) is reduced in intensity relative to the Co $L_{\eta,l}$ PFY and the transmission XA spectra. This fact could not be disentangled from the PFY data in Ref. [22] since it collects the total signal from $L_l$(3s → $2p_{3/2}$), $L_\eta$(3s → $2p_{1/2}$), $L_\alpha$(3d → $2p_{3/2}$) and $L_\beta$(3d → $2p_{1/2}$) channels. The observation of lowered pre-peak PFY intensities as well as dips in TFY for aqueous $Fe^{3+}$ and $Co^{2+}$ evidence a reduction of the fluorescence [20], e.g. due to electron delocalization into the solvent as discussed below.

This is supported by the previous partial-electron-yield study of Auger electrons [16], where a peak at 775 eV was reduced relative to the transmission spectrum. Hence, this mechanism is expected to decrease the non-radiative Auger-rate as well. Both, optical effect and fluorescence reduction mentioned above result in a disappearance of the first peak at the $L_3$-edge TFY of $Co^{2+}$.

The theoretical XA spectra are in fairly good agreement with transmission and PFY spectra for the investigated ions what validates the employed method (cf. Figures 1-3 and S4). In fact the present agreement seems to be superior to the level reached for $Cr^{3+}$ in Ref. [22] employing a similar scheme. The analysis of the present results showed that mixing of states having different multiplicities is important to reproduce the experimental spectrum. In the $Co^{2+}$ and $Fe^{2+}$ spectra, the most intense transitions correspond to strongly allowed 2p→3d($e_g$) $\Delta S = 0$ excitations. Remarkably, for $Fe^{3+}$ the XA spectrum is formed mainly by spin-forbidden sextet-quartet transitions though the most intense peaks are due to allowed sextet-sextet ones.

Most of the electronic states included in the calculations possess pronounced multi-configurational character and are additionally mixed due to spin-orbit coupling. On the one hand, this emphasizes the importance of the *n*-particle character of the RASSCF wave function and gives indication for a potential failure of TDDFT if applied to such problems. On the other hand, it hinders analysis and assignment of individual bands in terms of simple transitions between single particle MOs. However, for the low-energy side of the $L_3$-edge the dominant character could be deduced quite clearly. Since this spectral region is the one that shows the most pronounced distortions in the TFY spectra, the following discussion will focus on establishing a relation between these distortions and the character of the underlying states. The latter will be classified as being either of $e_g$ or $t_{2g}$ character.

As illustrated in Figures 1b and S2, the low energy transitions correspond to predominant excitation of core-electrons to $t_{2g}$ orbitals. It should be emphasized that $t_{2g}$ orbitals themselves do not extensively mix with water orbitals. But, through many-body effects the states in this region have contributions from configurations with a notable fraction of excited-core electron occupation on $e_g$ orbitals mixed with solvent orbitals. If the core-electron is excited to $e_g$ orbitals the extent of delocalization is notably higher and is now due to one-electron orbital mixing. This implies even larger charge delocalization for $e_g$ dominated states than for $t_{2g}$ dominated ones. Note that the presence of explicit water is vital for this picture to hold. Although a related point charge calculation yields a similar yet shifted spectrum, it cannot reproduce the nature of the MOs (see Figure S3).

In light of Figures 1b and S2 the observed dips in TFY can be explained using the following simple two-state model. Let us assume that atomic $2p$ orbitals of a metal ion ($|2p\rangle$) are the initial states whereas final states after electronic relaxation ($t = \infty$), i.e. the actual radiative states, are the already mentioned non-interacting $|t_{2g}\rangle$ and $|e_g\rangle$ states, where the core-electron is excited to $t_{2g}$ or $e_g$ 3d-type orbitals of metal ion mixed with oxygen $|2p_O\rangle$ states:

$$|t_{2g}\rangle = C_{3d}^{t_{2g}}(t=\infty)\,|3d_{t_{2g}}\rangle + C_{2p_O}^{t_{2g}}(t=\infty)|2p_O\rangle$$

$$|e_g\rangle = C_{3d}^{e_g}(t=\infty)\,|3d_{e_g}\rangle + C_{2p_O}^{e_g}(t=\infty)\,|2p_O\rangle.$$

Note that for the cases considered we have $C_{2p_O}^{t_{2g}}(t=\infty) < C_{2p_O}^{e_g}(t=\infty)$. The excitation at time $t = 0$ occurs via the local atomic transition dipole operator

$$\hat{D} = d_{2p \to t_{2g}}|3d_{t_{2g}}\rangle\langle 2p| + d_{2p \to e_g}|3d_{e_g}\rangle\langle 2p| + h.c.$$

and the wave function after excitation at time $t = 0$ depending on the wavelength reads

$$|\Psi(t=0)\rangle = \hat{D}|2p\rangle = C_{3d}^{t_{2g}}(t=0)\,|3d_{t_{2g}}\rangle \ \ or\ \ C_{3d}^{e_g}(t=0)\,|3d_{e_g}\rangle,$$

where $C_{3d}^{t_{2g}}(t=0) \propto d_{2p \to t_{2g}}$ and $C_{3d}^{e_g}(t=0) \propto d_{2p \to e_g}$. The time evolution of this state vector is that of simple two level system with the Hamiltonian

$$H = \begin{pmatrix} E_{3d} & V \\ V & E_{2p_O} \end{pmatrix}.$$

Assuming that $E_{3d} - E_{2p_O}$ and $V$ differ not much for both types of d-orbitals, an essential point is the difference in the emission life times determined by the absolute square of $d_{2p \to t_{2g}} C_{3d}^{t_{2g}}(t)$ or $d_{2p \to e_g} C_{3d}^{e_g}(t)$. Here for the most intense transitions $d_{2p \to e_g}/d_{2p \to t_{2g}} \approx 10$ holds according to the RASSCF/RASSI-SO calculations. Two cases can be distinguished: a) The excited state has dominant $t_{2g}$ character and the life time is relatively long due to the smallness of $d_{2p \to t_{2g}}$. This means that the electron delocalization to water *can occur completely* and the fluorescence intensity decreases. b) The excited state has dominant $e_g$ character what implies shorter life times and *incomplete electron relaxation*. As a result fluorescence intensity is not much affected. Thus based on the above picture, the ultrafast electron dynamics and the associated fluorescence spectra depend on the initial state selected by the wavelength of excitation. It is important to emphasize that we consider electron delocalization to water rather than 'real' metal-to-water charge-transfer leading to an n+1 times charged metal ion and a solvated electron.

Returning to the experimental data and taking into account that a) except for pre-peak the TFY, PFY and iPFY spectra are similar and b) both $Fe^{2+}$ and $Fe^{3+}$ have the same concentration in water and the background effect for both cases is similar, the observation of reduced pre-peak intensities and dips shows the existence of the non-radiative relaxation at the L-edge of aqueous $Fe^{3+}$. This conclusion is supported by XA studies on the solvent K-edge (K-edge of waters' oxygen) [34], L-edge XA spectra of TM complexes [12], and photoelectron spectroscopy [16].

In case of aqueous $Fe^{3+}$ ion, there is an extended region 707.0-709.5 eV where the excited electron occupies dominantly $t_{2g}$ states (shaded area in Figure 1b) and the decrease in fluorescence intensity leads to dips below the background level. Note, that the same holds true for $[FeCl(H_2O)_5]^{2+}$ and $[FeCl_2(H_2O)_4]^{2+}$ which, together with $[Fe(H_2O)_6]^{3+}$, are the main species in solution under experimental conditions (see, Figs. S4 and S5). For $Fe^{2+}$ and $Co^{2+}$, $e_g$ occupation starts to dominate at the low energy edge and the fluorescence is less affected (see Fig. S2). If we now take into account that in case of $Fe^{2+}$ the orbital mixing is less pronounced ($C_{2p_O}^{t_{2g}}(t=\infty)$ and $C_{2p_O}^{e_g}(t=\infty)$ are calculated to be a factor of three smaller than for $Fe^{3+}$), this indeed supports the experimental observation where the dip in TFY is strongest for $Fe^{3+}$, weaker for $Co^{2+}$ and almost negligible for $Fe^{2+}$. Note, that $Fe^{2+}$ in this respect behaves similar to $Cr^{3+}$ studied recently [22]. Thus, the extent of delocalization of excited core-electron on solvent molecules through single-particle (orbital mixing) and many-particle (electron correlation) effects observed in theoretical calculations correlates with the extent of TFY distortion in the experiments (this cannot be captured by a point charge model, see Fig. S3). The differences in radiative life times for different core-excited states explain why the electron delocalization is state-dependent and spectral distortions in PFY and TFY relative to the true absorption cross-section are observed only for the pre-peak region.

Summarizing, the combination of *ab initio* multi-configurational calculations (RASSCF/RASSI-SO) with recently developed high-resolution PFY and iPFY for the study of L-edge XA spectra of TM aqueous ions demonstrated that beside the well-known background effect (X-ray optical effect), electron delocalization across the metal-solvent interface can be, depending on the metal species, responsible for the TFY and PFY spectral distortions. For $Fe^{3+}$ and partially also for $Co^{2+}$ this behavior depends on the nature of the excited state and the ratio between its radiative life time and electronic relaxation time. Here the fluorescence dips correspond to the predominant $t_{2g}$ occupation of the core-electron. Although the $e_g$ states of the d-orbitals are more strongly mixed with the solvent molecular orbitals, because of longer life times, $t_{2g}$ states lead to a more pronounced electron delocalization into the water solvation shell.

This work was supported by the Helmholtz-Gemeinschaft via the VH-NG-635 grant (E.F.A.) and the European Research Council starting-grant no. 279344 (E.F.A.).

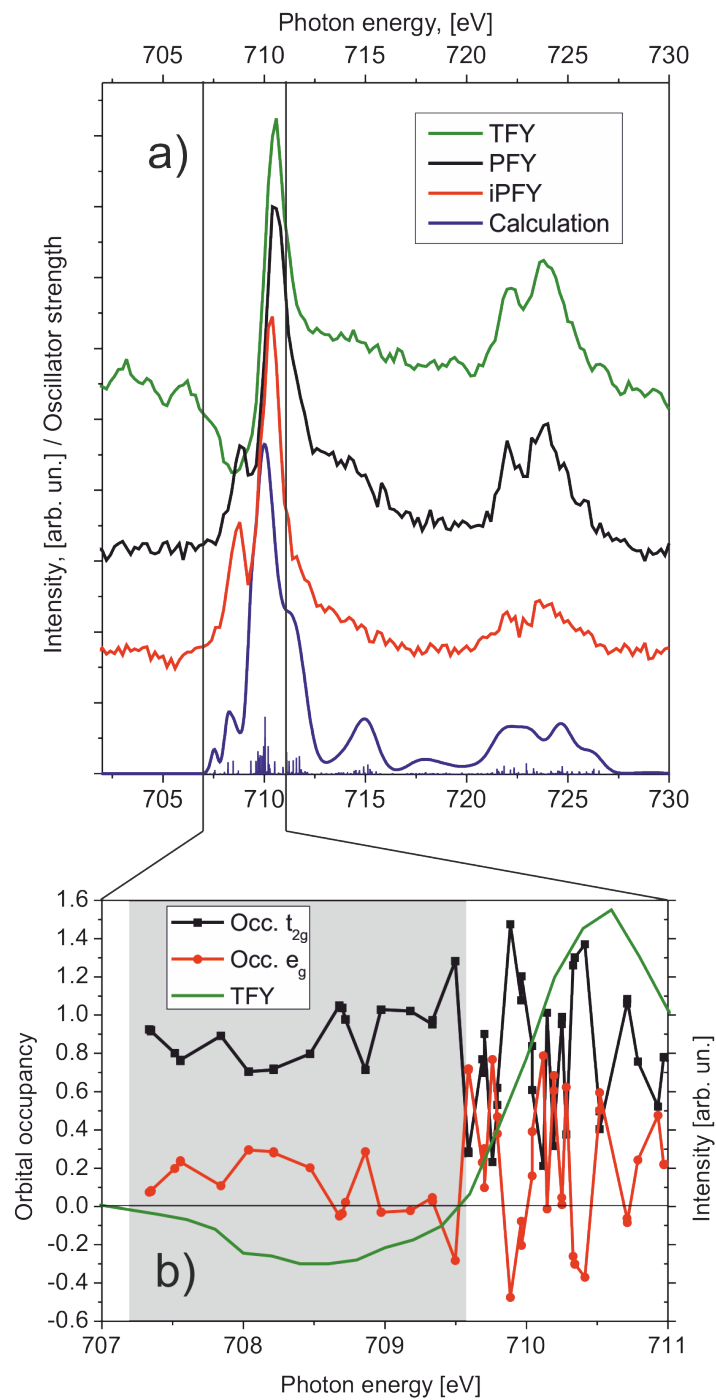

Figure 1 (Color). a) Experimental TFY, iron PFY, oxygen iPFY spectra of 1M $FeCl_3$ solution in water as compared to theoretical XA spectrum for $[Fe(H_2O)_6]^{3+}$. b) Occupation numbers for $t_{2g}$ and $e_g$ 3d orbitals of metal ion correlated to the dip in TFY spectrum. The shaded area marks the region where a core-electron is predominantly excited to $t_{2g}$ orbitals and the maximal distortion of the TFY occurs.

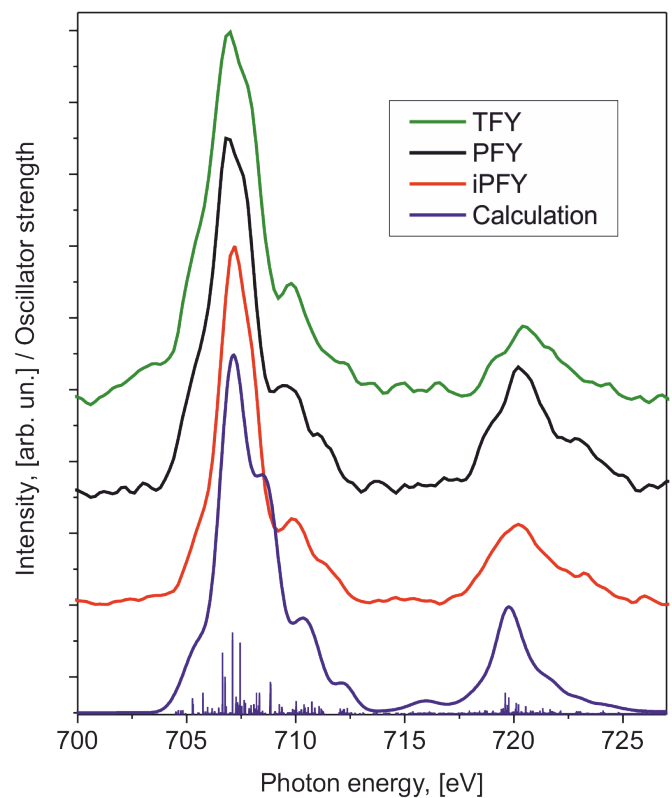

Figure 2. Experimental TFY, iron PFY, oxygen iPFY spectra of 1M $FeCl_2$ solution in water [19] as compared to theoretical XA spectrum for $[Fe(H_2O)_6]^{2+}$.

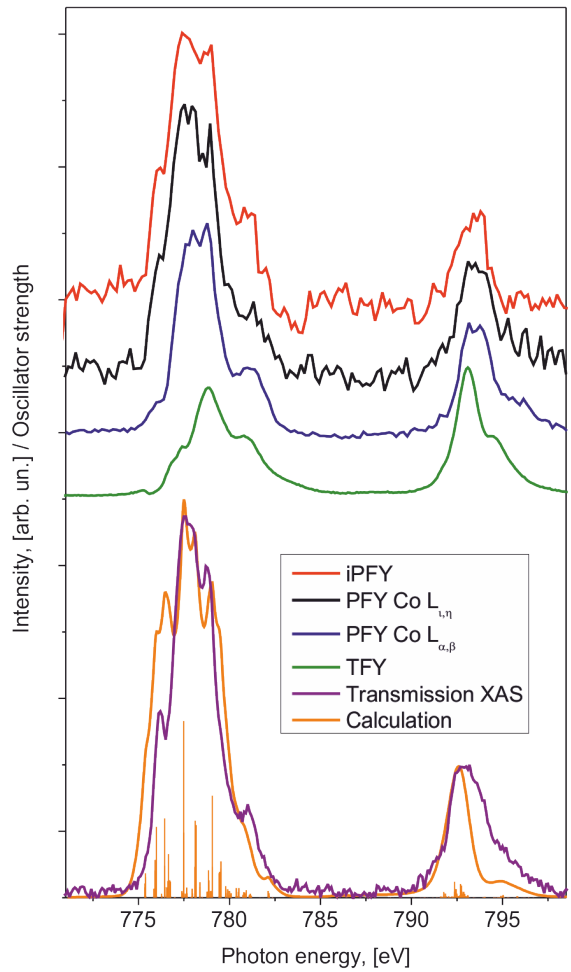

Figure 3. Experimental oxygen iPFY, cobalt PFY $L_{\iota,\eta}$ and $L_{\alpha,\beta}$, TFY, and transmission XAS spectra of 1M $CoCl_2$ solution in water [20] as compared to theoretical XA spectrum for $[Co(H_2O)_6]^{2+}$.

# Supplementary Material
# State-Dependent Electron Delocalization Dynamics at the Solute-Solvent Interface: Soft X-ray Absorption Spectroscopy and Ab Initio Calculations

Sergey I. Bokarev[1], Marcus Dantz[2,3], Edlira Suljoti[2,3], Oliver Kühn[1], Emad F. Aziz[2,3]

[1]Institut für Physik, Universität Rostock, Universitätsplatz 3, D-18055 Rostock, Germany
[2]Joint Ultrafast Dynamics Lab in Solutions and at Interfaces (JULiq), Helmholtz-Zentrum Berlin für Materialien und Energie, Albert-Einstein-Strasse 15, D-12489 Berlin, Germany
[3]Freie Universität Berlin, FB Physik, Arnimallee 14, D-14195 Berlin, Germany

**Computational details.** In the theoretical simulations, cluster models of aqueous ions with their first solvation shell were used, i.e. $[M(H_2O)_6]^{n+}$, where M = $Co^{2+}$, $Fe^{2+}$, and $Fe^{3+}$ as well as $[FeCl(H_2O)_5]^{2+}$ and $[FeCl_2(H_2O)_4]^{+}$ which are together with $[Fe(H_2O)_6]^{3+}$ the main species in $FeCl_3$ aqueous solution under experimental conditions. The ground state geometries of the high-spin electronic states were optimized at the Møller-Plesset perturbation theoretical level (MP2/cc-pVTZ) for $Co^{2+}$ and $Fe^{3+}$ using the Gaussian 09 program package [1]; for $Fe^{2+}$ the geometry from its multi-reference analogue CASPT2(10,12)/ANO-RCC reported in [2] was adopted. Jahn-Teller distortion leads to $D_{2h}$ symmetry of both $[Co(H_2O)_6]^{2+}$ and $[Fe(H_2O)_6]^{2+}$, whereas $[Fe(H_2O)_6]^{3+}$ retained octahedral symmetry of the surrounding oxygen atoms resulting in a $T_h$ point group symmetry. To simplify the discussion all 3d orbitals of the three ions are treated as being $t_{2g}$ and $e_g$ like MOs. XA spectra were calculated on the RASSCF level with the relativistic ANO-RCC-VTZ basis set for Co, Fe, Cl and O and ANO-RCC-VDZ for H atoms [3, 4]. These calculations implied no symmetry. The active space consisted of nine orbitals (Figure S1b): three 2p (RAS1 space), five 3d (RAS2 space), and 4s (RAS3) orbitals of metal ions including the dipole-allowed 2p→3d and 2p→4s transitions as well as shake up 3d→3d excitations. One hole and one particle were allowed in the RAS1 and RAS3 subspaces, respectively; the RAS2 subspace was treated with full configuration interaction. The orbitals were first optimized in a state-averaged complete active space calculation (active space without 2p orbitals) for the lowest five valence high-spin states and then all occupied orbitals except 2p and 3d were kept frozen. The spin-orbit coupling was treated within the RASSI-SO method [5] including directly interacting states with $\Delta S = 0, \pm 1$. In addition, the number of states included here appeared to be an essential issue and to minimize the error due to the choice of the states all possible one-electron core-excitations were taken into account (see Table S1). Scalar relativistic effects were considered within the Douglas-Kroll-Hess approach [6, 7]. Only excitations from the lowest spin-component of the high-spin ground state were taken into account. Theoretical spectra were broadened with a pseudo-Voigt lineshape function. RASSCF/RASSI calculations were performed with MOLCAS 7.6 program suite [8].

Note, for the purpose of comparison the simulated spectra are shifted to the blue with respect to experiment by 1.1 and 1.8 eV, for $Co^{2+}$ and $Fe^{2+}$, respectively, while $Fe^{3+}$ spectra (with and without chlorines) remained unshifted. These values are quite small if compared to TDDFT shifts for K-edge spectra [9, 10]. In general, this shift can be attributed to insufficient flexibility (minimal basis quality) of the ANO-RCC bases in the core-region and frozen relaxation of a fraction of valence orbitals upon core-hole formation.

For the qualitative interpretation of fluorescence dips we have developed a simple two state model as detailed in the main text. Note that this model does account for a selection of most important states of the solute only. This implies in particular that effects due to the solvent background are not treated, because they cannot be quantified on the present level of ab initio calculations.

Table S1. Number of spin-free (SF) and spin-orbit (SO) states used for RASSI-SO calculation of core-excitation spectrum.

|  | $[Co(H_2O)_6]^{2+}$ | $[Fe(H_2O)_6]^{2+}$ | $[Fe(H_2O)_6]^{3+}$ |
|---|---|---|---|
| SF $\Delta S = 0$ | Quartets (S=3/2) 270 | Quintets (S=2) 225 | Sextets (S=5/2) 99 |
| SF $\Delta S = -1$ | Doublets (S=1/2) 480 | Triplets (S=1) 729 | Quartets (S=3/2) 596 |
| SF $\Delta S = +1$ | Sextets (S=5/2) 35 | Septets (S=3) 16 | Octets (S=7/2) 3 |
| SO total | 2250 | 3424 | 3002 |
| SO core-excited | 1710 | 2835 | 2670 |

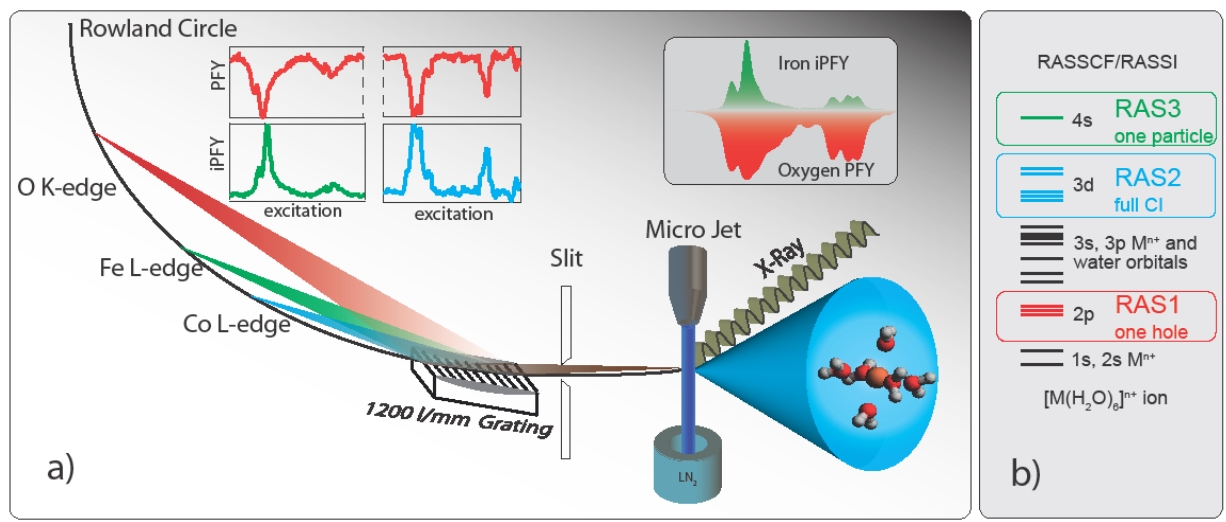

Figure S1. a) Experimental setup used for measurement of PFY and iPFY spectra. b) Orbital active space consisting of 9 MOs subdivided into three subspaces used for RASSCF calculations of $[M(H_2O)_6]^{n+}$ XAS.

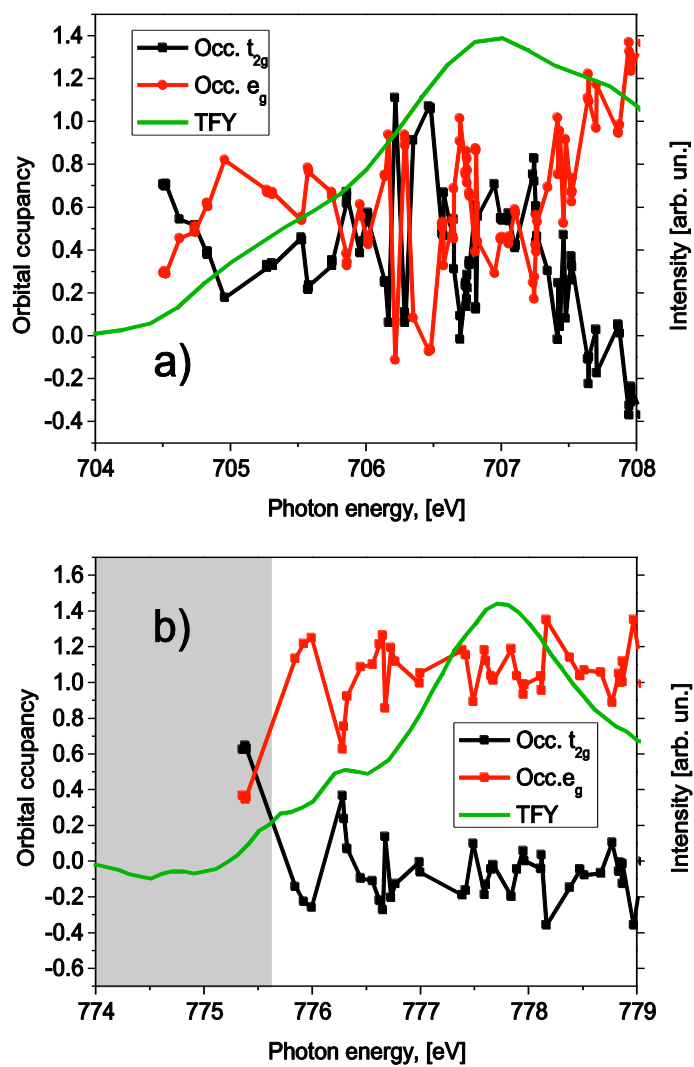

Figure S2. Occupation numbers for $t_{2g}$ and $e_g$ 3d orbitals of a) aqueous $Fe^{2+}$ and b) aqueous $Co^{2+}$ overlaid by TFY spectrum. Shaded area marks the region where the distortion of TFY occurs.

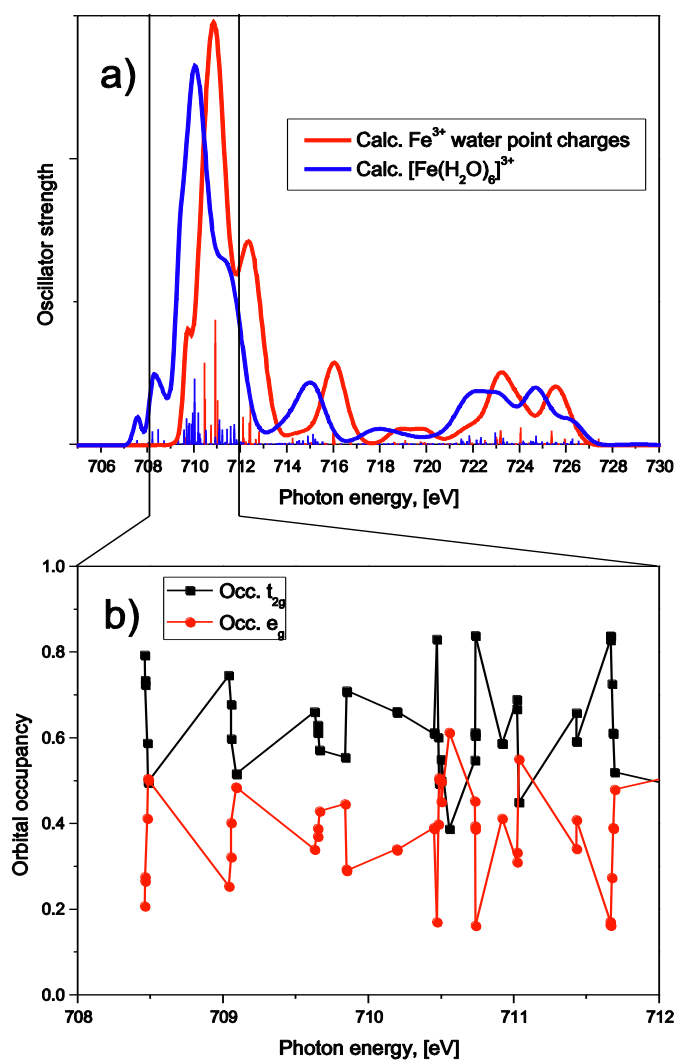

Figure S3. a) Results of theoretical calculations of $Fe^{3+}$ ion with point charges modeling water molecules and with explicit inclusion of water allowing electron delocalization. This delocalization leads to the shift of the spectrum as a whole with respect to $Fe^{3+}$ with point charges, what can be understood in terms of particle in a box model. In addition, the intensity of the $L_3$-edge decreases going from charges to explicit waters what illustrates the lowered d-contribution in both $t_{2g}$ and $e_g$ orbitals after relaxation (see main text). b) Occupation numbers for $t_{2g}$ and $e_g$ orbitals of $Fe^{3+}$ ion with point charges. Note that for the low energy spectral region the occupancies are less systematic than for explicit waters (see Figure 1b).

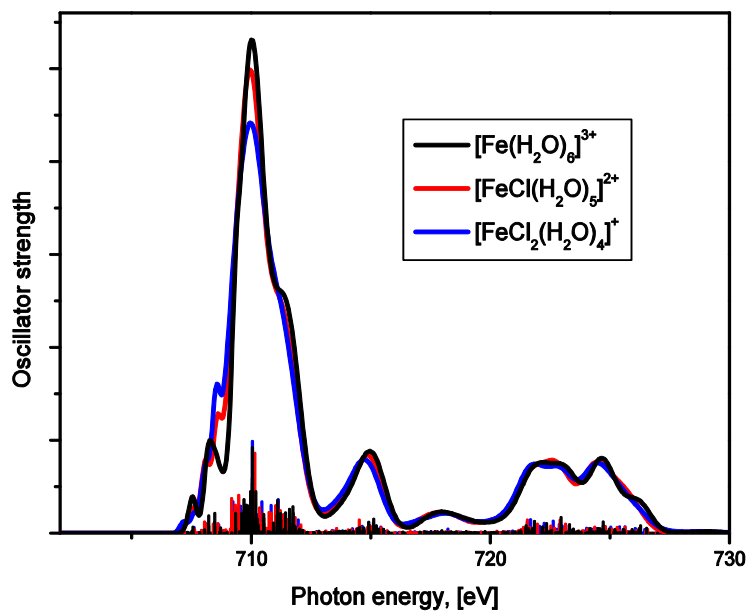

Figure S4. Results of theoretical calculations of $[Fe(H_2O)_6]^{3+}$, $[FeCl(H_2O)_5]^{2+}$, $[FeCl_2(H_2O)_4]^+$ ions which are possible species in $FeCl_3$ aqueous solution under experimental conditions used. The intensity of the main $L_3$-peak decreases upon association with $Cl^-$ ions, whereas the intensity of pre-peak rises. The rest of the spectrum stays almost unchanged. The theoretical spectrum of $[FeCl_2(H_2O)_4]^+$ is in better agreement with experimental iPFY spectrum (Fig. 1) due to higher pre-peak intensity.

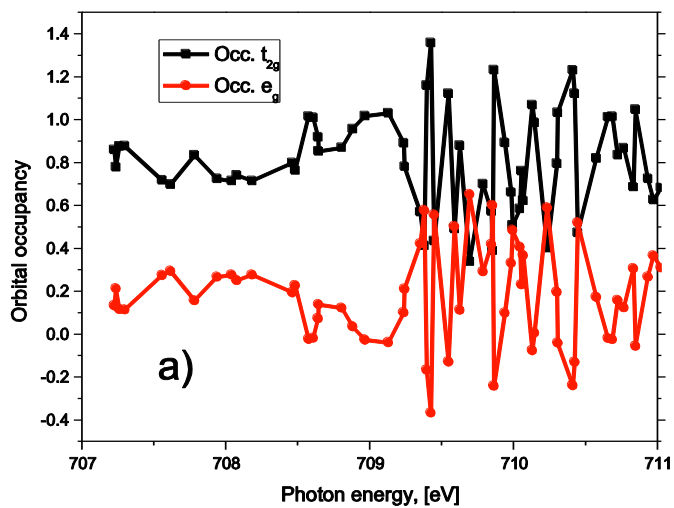

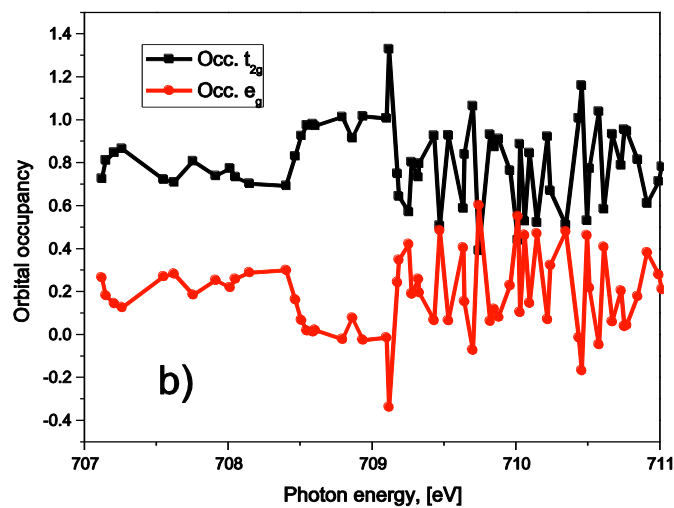

Figure S5. Occupation numbers for $t_{2g}$ and $e_g$ 3d orbitals ($O_h$ point group symmetry labels are used) for a) $[FeCl(H_2O)_5]^{2+}$ and b) $[FeCl_2(H_2O)_4]^+$. Calculations show that association with $Cl^-$ ion does not much change the nature of the core-excited states if compared to $[Fe(H_2O)_6]^{3+}$ (see Figure 1b).